\documentclass[10pt, conference, compsocconf]{IEEEtran}

\usepackage{epsfig}
\usepackage{subfigure}
\usepackage{graphicx}
\usepackage{xcolor}
\usepackage[normalem]{ulem}
\usepackage{comment}
\usepackage{multirow}
\usepackage{lscape}
\usepackage{listings}
\usepackage{tabularx}
\usepackage{lineno,hyperref}
\usepackage{blindtext}

\begin{document}

\title{Microservices: Migration of \\ a Mission Critical System}



\author{
\IEEEauthorblockN{Nicola Dragoni\IEEEauthorrefmark{1}, 
Schahram Dustdar\IEEEauthorrefmark{2},
Stephan T. Larsen\IEEEauthorrefmark{3}, Manuel Mazzara\IEEEauthorrefmark{4}}
\IEEEauthorblockA{\IEEEauthorrefmark{1}Technical University of Denmark and \"Orebro University, Sweden\\ ndra@dtu.dk}
\IEEEauthorblockA{\IEEEauthorrefmark{2} TU Wien \\ dustdar@dsg.tuwien.ac.at}
 \IEEEauthorblockA{\IEEEauthorrefmark{3}Danske Bank, Denmark \\ stephantl@gmail.com}
    \IEEEauthorblockA{\IEEEauthorrefmark{4}Innopolis University, Russia \\ m.mazzara@innopolis.ru}
}

\maketitle

\begin{abstract}
The microservices paradigm aims at changing the way in which software is perceived, conceived and designed. One of the foundational characteristics of this new promising paradigm, compared for instance to monolithic architectures, is scalability. In this paper, we present a real world case study in order to demonstrate how scalability is positively affected by re-implementing a monolithic architecture into microservices. The case study is based on the \textit{FX Core} system, a mission critical system of Danske Bank, the largest bank in Denmark and one of the leading financial institutions in Northern Europe. 
\end{abstract}









\section{Introduction}
\label{section:intro}

The history of software architectures have been characterized in the last few decades by a progressive shift towards distribution, modularization and loose coupling, with the purpose of increasing code reuse and robustness \cite{AlmeidaALGM04}, ultimately a necessity dictated by the need of increasing software quality, not only in safety and financial-critical applications, but also in more common off-the-shelf software packages. 

The latest step in this process seems to be named under \textit{microservice architecture}, which is a style inspired by service-oriented computing that has recently started gaining popularity and that promises to change the way in which software is perceived, conceived and designed \cite{Dragoni2017}. In service-oriented architectures \cite{mackenzie2006}, the emphasis was mostly on cross-boundaries inter-organization technology-agnostic communication, and on orchestration of business processes \cite{YanMCU07}. The research community dedicated major attention and effort on foundational aspects, such as correctness and verifiability of service composition \cite{Mazzara:phd}. Little effort was spent on defining and bounding the nature of the internal logic of services or on scalability and maintainability issues, concerns that appear to be of major importance for modern organizations. New programming languages based on the new paradigm are emerging in recent years\cite{MGZ14}, and allow to describe computation from a data-driven instead of process-driven perspective \cite{Safina2016}.

The shift towards microservices is a sensitive matter these days, seeing several companies involved in a major refactoring of their back-end systems to accommodate the easiness of the new paradigm. This is the case, for example, of the system and the institution considered in this paper, i.e., the \textit{FX Core} of Danske Bank. Other companies instead just start their business model developing software following the microservice paradigm since day one. We are in the middle of a major change in the view in which software is perceived and intended, and in the way in which capabilities are organized into components, and industrial systems are conceived.

The microservices architecture \cite{Dragoni2017} is built on very simple principles:

\begin{itemize}
\item \emph{Bounded Context:} first introduced in \cite{evans2004domain}, this concept captures one of the key properties of microservice architecture: focus on business capabilities. Related functionalities are combined into a single business capability which is then implemented as a service.

\item \emph{Size:} this represents a crucial concept for microservices and
brings major benefits in terms of service maintainability
and extendability. Idiomatic use of a microservice
architecture suggests that if a service is too large, it
should be refined into two or more services, thus preserving
granularity and maintaining focus on providing
only a single business capability.

\item\emph{Independency:} this concepts encourages loose coupling and high cohesion by stating that each service in microservice architectures is operationally independent from others, and the only form of communication between services is through their published interfaces.
\end{itemize}

Implementing a system in the microservice architectural style enables it to handle scale almost out of the box. Many of the techniques and principles used to implement a microservice architectures provide an architecture with characteristics that are beneficial to scalability. The main microservice characteristics and how they naturally contribute to system scalability have firstly been introduced in \cite{Dragoni2017b}, although no practical case study has been considered in that contribution.

\subsection*{Contribution of the Paper} 
In this paper, we consider a real world case study concerning the migration of a mission critical system from an existing monolithic architecture to microservices, i.e., the \textit{FX Core} system of Danske Bank, the largest bank in Denmark and one of the leading financial institutions in northern Europe. The contributions of the paper are threefold. First, we highlight the key technical aspects that neeed to be considered for full system scalability in the microservice context. Second, we show how a real world monolithic architecture can be converted into a microservice one, and we highlight the resulting benefits of this migration. Finally, we study in detail how scalability has positively been affected by this paradigm transition. To the best of our knowledge, this is the first real world case study of migrating a monolithic system to a microservice one with focus on scalability.\\

\noindent\emph{Outline of the Paper.} The paper is structured as follows: Section~\ref{sec:enablingscalability} discusses the technical aspects that are necessary to consider to exploiting the full scalability potential of microservices. In Section~\ref{sec:danskebank} the Danske Bank \textit{FX Core} system functionalities are described, as well as its general structure. The legacy monolithic architecture is then presented in Section~\ref{sec:legacysystem}, while the 
proposed microservice-based one appears in  Section~\ref{sec:fxcoremsa}. The comparison between the two architectures is detailed in Section~\ref{sec:comparisondbmonomsa}. Section~\ref{sec:conclusion} concludes the paper.
\section{Microservice Architecture and Scalability}
\label{sec:enablingscalability}

Proponents of the Microservice architecture claim, that per se this style increases system scalability. However, it is necessary to pay particular attention to certain technical features, and spend extra effort on them in order to fully enable the potential. This section will cover all the aspects that need to be taken care of to enable full scalability. i.e., \textit{automation, orchestration, service discovery, load balancing, and clustering}.

\subsection{Automation}

In monolithic architectures, at times it is possible to manually manage the system and the hosts on which it is running. However, as soon as the system scales, the number of hosts may increase leading to a hard-to-maintain system. This complexity applies to microservice architecture too, where services are scattered across multiple hosts, with each one running multiple services. Manually managing a microservice architecture would result in an enormous time overhead, since deployment, configuration, and maintenance now extends to each and every service instance and host. Every time a new service or host is introduced, the system will require an increasing amount of time for manual management. When standard management activities, such as builds, tests, deployment, configuration, host provisioning and relocation of services are automated, the introduction of new services does not imply a management overhead. Only maintenance of scripts is required, and developers are expected to manage all the system via automation. The bottom line is automation of growth-sensitive tasks, in order to contain the time overhead.

\subsection{Orchestration}\label{subsec:enablingscalabilityorchestration}


In Microservice architectures orchestration is necessary for managing service containers and infrastructure. Without an orchestration system, engineers should develop and maintain a number of features which are necessary to run a system at large scale. Open source orchestration systems such as Google's \textit{Kubernetes}~\cite{kubernetes2016}, Mesosphere's \textit{Marathon}~\cite{marathon2016} and Docker's built-in \textit{Swarm Mode}~\cite{dockerswarm2016} all provide a number of features which are necessary to achieve scalability, such as \textit{service discovery}, \textit{load balancing} and \textit{cluster management}. Orchestration systems also handle replication of services and distribution of replicas across the nodes.

\subsection{Service Discovery}
\label{subsec:enablingscalabilityservicediscovery}

A widespread diffusion of the traditional Service-Oriented architecture has been prevented by fundamental shortcomings related to service discovery and, in particular, dynamic binding and invocation \cite{Michlmayr2007}. Microservices and orchestration tools are trying to overcome these issues.

Microservice architectures consist of many services, and a mechanism has to be deployed to keep track of which service instances are running, and how to reach them. This is typically done with a \textit{service discovery} tool, either a separate service such as \textit{Consul}~\cite{consul2016}, or as part of the aforementioned orchestration tools. Service discovery provides more than simple DNS lookups, it also provides mechanisms for health-checking, ensuring that the services it resolves names to, are actually alive.

Service discovery is a must in microservice architectures, since services do not have static IP addresses and require a mapping from a hostname. Service discovery can make use of \textit{locality}, resolving hostnames to the service instance that is closest to the requester, hereby achieving \textit{geographical scalability}. Service discovery also creates the illusion of interacting with a single service, although a sequence of requests actually might be handled by multiple service replicas.


\subsection{Load Balancing}
\label{subsec:enablingscalabilityloadbalancing}
\textit{Load balancing} is a key part of service discovery, necessary to ensure that all load is not sent to a single service. Load balancing distributes load across service replicas in a variety of ways. By use of DNS mechanisms, which can resolve hostname, lookups to a different replica IP each time, by using a dedicated server, i.e. a \textit{dispatcher} to hand out replica IP's based on some scheduling algorithm or by letting the clients themselves decide which replica to connect to. This is typically implemented as part of service discovery and orchestration. If the services integrate via a central messaging system,  these can also be used to distribute messages and events to a different replica each time, hereby distributing load.

\subsection{Clustering}\label{subsec:enablingscalabilityclustering}
Microservice architectures can be deployed on a single host. However, this would not contribute to scalability. To enable full scalability, deployment has to happen on a cluster of hosts. Clustering enables a system to utilize multiple hosts resources as a single system. It also enables elasticity in the form of expansion with additional hosts when needed and decrease of hosts when not. This may be achieved without clustering, but it would require the entire system to run on each host, like vertically scaled monolithic architectures. 

These computing clusters can be configured and run with a variety of tools, but if containerization is used it will typically be part of the orchestration
tooling. These tools will ensure that services and replicas are spread across
the cluster and are able to reach each other, enabling higher availability,
increased resilience and better load scalability.

Running services in a cluster, requires them to either run actively in
parallel or, in the case of infrastructure and data storage, to use clustering mechanisms in order to collaborate. These clustering mechanisms differ depending on their requirements to performance, consistency, availability etc., but are typically included in scalable messaging systems, such as \textit{RabbitMQ} \cite{rabbitmq2016}, and databases such as \textit{Redis} \cite{redis2016},
\section{Danske Bank \textit{FX Core} system}
\label{sec:danskebank}
The Danske Banks \textit{FX Core} system is a paradigmatical case study to demonstrate how to effectively migrate from a monolithic to a microservice architecture, and how this affects scalability. The documentation of the original system architecture was sparse and the vast majority of technical  details have been obtained by direct conversations, interviews and discussions with the \textit{FX Core} team, and by manually inspecting the source code. This was a lengthy process given the complexity of the monolithic architecture . The outccome of this process is reported in this paper, where we describe the system in terms of responsibilities and organization. All confidential information, such as concrete names of rotocols, external providers and specific services,has been withheld in order for the results to be published.

\subsection{Foreign Exchange}
\textit{Foreign Exchange}, often abbreviated as \textit{forex} or \textit{FX}, is the exchange of currencies, i.e.\ the conversion from one currency to another. Exchange of currencies is of interest to both private individuals, corporations, financial institutions and governments. \textit{FX} encompasses everything from private transactions performed in foreign countries  (e.g.\ Internet shopping from abroad and use of credit cards while traveling) to corporations moving their financial assets from one currency to another and exporting or importing products to and from foreign markets.\textit{FX}  has grown with globalization and it is now globally the largest financial market in the world, averaging a daily transaction volume of roughly 5 trillion dollars. This results in some transactions reaching the hundred millions of dollars. Unlike the stock exchange, there is no centralized market, instead \textit{FX} is decentralized and done \textit{over-the-counter (OTC)}, i.e.\ traders negotiate prices and trade directly between each other. Traders are typically the largest multinational banks, trading on behalf of their customers or themselves. Additionally, due to the decentralized and global nature of \textit{FX} , the market is open 24 hours a day, five days a week~\cite{investopedia2017web}.

\subsection{FX IT}
\begin{figure}[!htbp]
    \centering
    \includegraphics[width=0.5\textwidth]{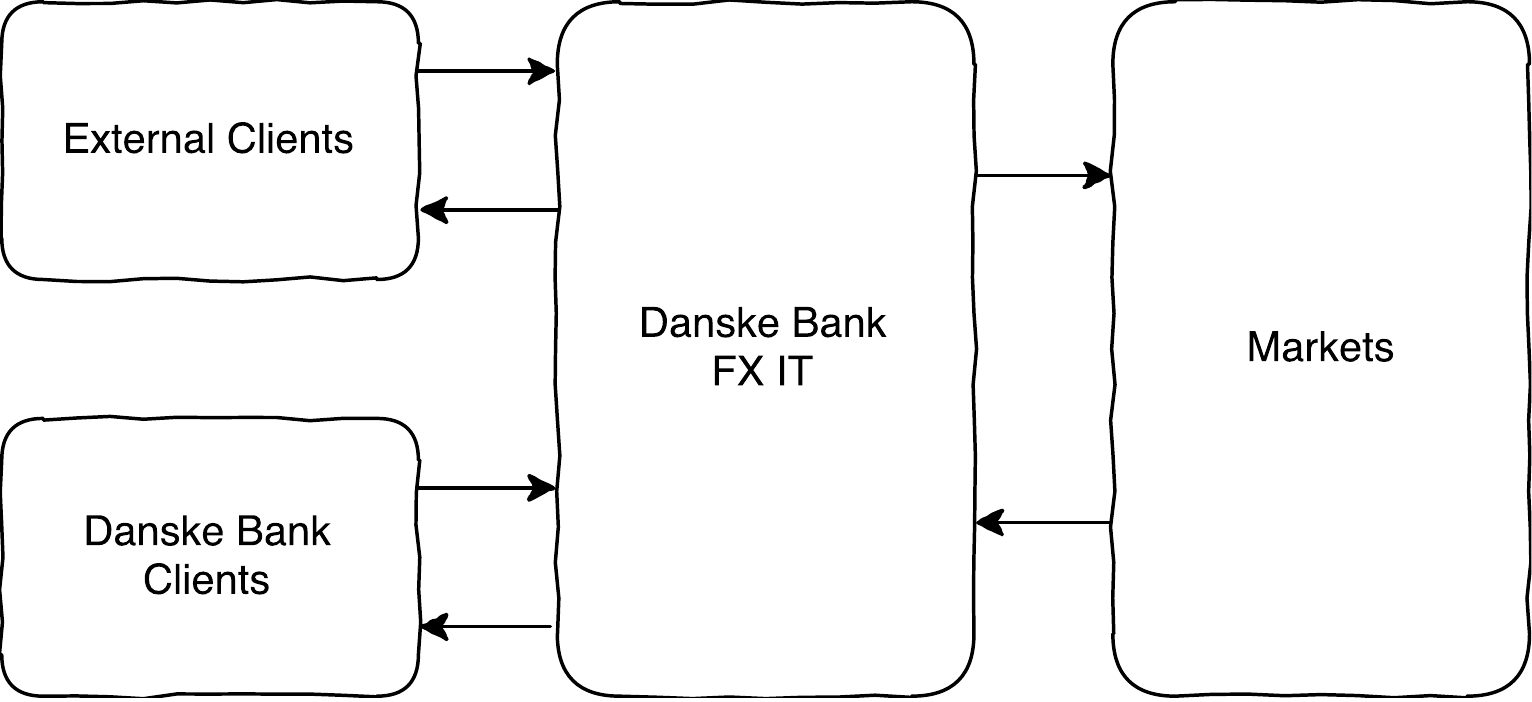}
    \caption{FX IT handles both price streaming and requests for trades and line-checks from global markets, e.g. other banks, pension funds and large corporations. Prices of currency pairs are streamed to FX IT, which then calculates prices of specific trades, before streaming them to external and internal clients. The clients can then request FX IT for trades or line-checks on the prices they have received. These clients are usually used by Danske Banks internal traders and external customers. Additionally, trade and line-check requests can also be received from the markets, when banks wish to exchange currencies directly. \textit{FX Core} is part of \textit{FX IT}, but handles tasks associated with trades and line-checks, thus not handling any of the price streaming and stream processing.}
\label{fig:fxit}
\end{figure}

The \textit{FX IT} (Fig~\ref{fig:fxit}) system is part of the banks \textit{Corporates and Institutions (C\&I)} department and handles price streaming, trades, line-checks and
associated tasks, such as analytics and post-trade management. \textit{FX IT} acts as
a gateway between the international markets and Danske Banks clients, including their own traders. C\&I's clients are mainly large financial institutions and large multi-national corporations. They continuously process streams of currency pair prices from the markets on which they calculate margins to reduce risk, especially important on \textit{swaps} and \textit{forwards}, before streaming final prices to clients. Clients can then act on a price by registering a trade or check if they have the required collateral with line-checks.

\subsection{FX Core}

The \textit{FX Core} system is part of \textit{FX IT} and handles trades and line-checks. This includes registration, validation and post-trade management. Below is a brief description of the two main responsibilities \textit{FX Core} has, i.e.\ trades and line-checks.\\

\noindent \emph{Trades} are received from both Danske Banks clients and \textit{external providers}, i.e.\ external clients and markets. The trade is then validated and line-checked, before being registered. Validations include authorization of clients, checks if trade is located on banking dates and if the currency pair price is valid. Depending on the type of trade, the trade is either done immediately, i.e.\ a \textit{spot trade}, or registered in the system as a contract for future execution, i.e.\ \textit{swaps} and \textit{forwards}. When the trade is executed it involves moving the financial assets between banking books, i.e.\ from one account to another. After a trade has been registered a number of actions can be executed on it, e.g.\ multiple trades can be joined to ease administration or be split into smaller trades to reduce margins, \textit{forward} and \textit{swap} contracts can be extended or pre-settled and trades can be corrected or deleted by internal clients. Additionally, the system can also run batch jobs in order to balance books between departments or to analyze trades, to e.g.\ detect fraudulent behavior such as money laundering.\\
 
\noindent \emph{LineChecks} are used to check whether a client has the financial collateral to perform a trade and how a trade will affect said collateral, also called their \textit{Line}. This collateral can be a multitude of financial assets, e.g.\ stocks, bonds or cash. Line-checks are always executed as part of a trade, but is also run separately, so Danske Banks traders can ensure that their customers are capable of requested trades.
\section{FX Core Monolith}
\label{sec:legacysystem}
In order to evaluate the benefits of implementing the \textit{FX Core} system as a microservice architecture, the monolithic architecture it replaces will be covered in this section. This includes an overview of the architecture and how it copes with scale, including the applied scalability techniques and the characteristics achieved. This section will also cover problems with the monolithic architecture which are not related to scale, but that still motivated the redesign of the system.

\subsection{Architecture}
\begin{figure}[!htbp]
    \centering
    \includegraphics[width=0.5\textwidth]{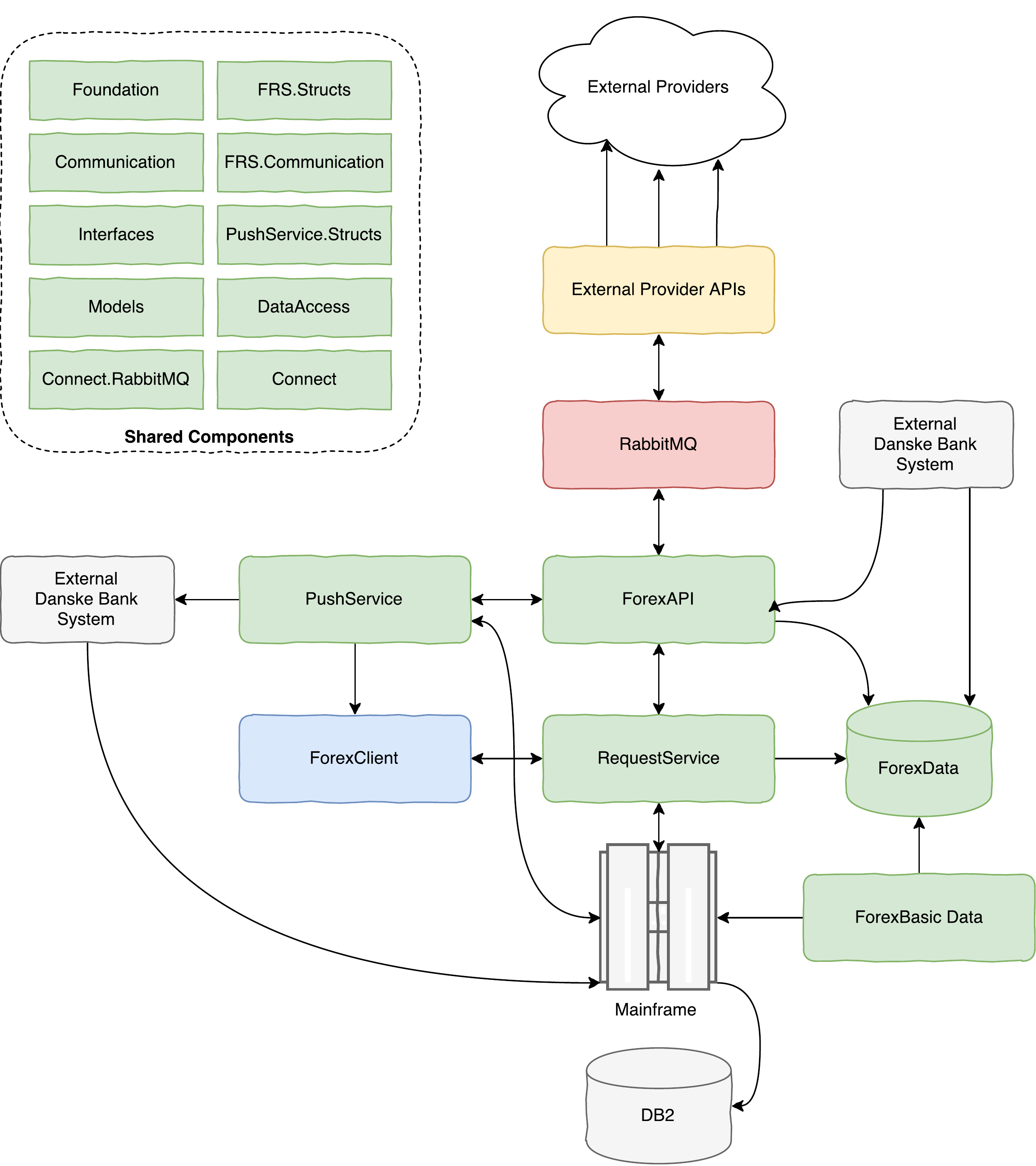}
    \caption{Danske Banks monolithic architecture. Red services are
    infrastructure services, green are part of the monolith, blue is the client, yellow are external provider APIs and grey are external Danske Bank systems.The components of the system integrate directly with each other,     resulting in many different communication technologies and high coupling.    The external provider APIs are part of the monolith and consist of multiple services, with each one connecting to a different provider. Their names have
been excluded due to confidentiality. The shared components are used across almost all services and are also internally dependent on each other. The ForexData database is one big monolithic MS SQL database, shared amongst many of the monolithic components and also accessed by external systems}
\label{fig:danskebank_monolith}
\end{figure}

Danske Banks monolithic system was in part already service-based, as it can be seen in Figure~\ref{fig:danskebank_monolith}. The system copes with scale in a variety of different ways. The services are  deployable individually, and are actually already replicated and deployed across a cluster. The system also utilizes APIs as interfaces for clients to interact with the services of the system, and a messaging system to delegate received requests from external providers. At a first sight, it  looks like an ideal and scalable solution. However, Danske Bank has experienced severe challenges when trying to rapidly develop the system and deploying consistent changes, and in general in handling system complexity. We will describe here systems components, how they integrate and how they are deployed. 

\subsubsection{System Components}
\label{subsubsec:fxcoremonolithsystemcomponents}

The monolithic architecture, shown in Figure~\ref{fig:danskebank_monolith}, is componentized in a variety of ways. The system utilizes both services, shared software libraries and thick desktop clients.
This section will briefly cover each of these components, their type and their responsibilities, in order to give an idea of how functionalities and data is distributed across the system. The thick clients will not be covered, as they are not going to be replaced by the microservice architecture, but simply be updated to interface with it.

\paragraph{External APIs}
The external APIs integrate with external providers of trades and line-checks, and due to confidentiality reasons will not be mentioned explicitly. All the APIs have TCP sockets open for communication with the external providers. The APIs receive requests for trade
and line-checks from the different providers, each with their own trading
protocol and feeds them into the system via the messaging queues in
\textit{RabbitMQ} \cite{rabbitmq2016}. The protocols are not translated by the APIs, so requests are simply fed to the system \textit{as is} wrapped in messages. The APIs are two-way, to notify the external providers with status of their line-checks and trades. The APIs receive responses via \textit{RabbitMQ} as well. All the external APIs function as services running in their own processes.

\paragraph{ForexAPI}
The \textit{ForexAPI} receives all requests for trades and line-checks from the external APIs through \textit{RabbitMQ}. It translates the proprietary protocols from the APIs to a uniform local format, and the other way around when responding. This results in a lot of translation
logic which ideally, due to the practice of separating concerns, should have been provided by the external APIs themselves, before entering the system. Some of the translation logic also resides in \textit{RequestService}. Not all requests for trades and line-checks are
received through \textit{RabbitMQ}, since multiple external Danske Bank systems can access the \textit{ForexAPI} directly via Remote Procedure Call (\textit{RPC}). Some integration with \textit{RequestService} is done via RPC and some through \textit{ForexData}, which is shared between the two. \textit{ForexAPI} also provides interfaces to external clients and users for several system functionalities, which it either handles itself or mediates to \textit{RequestService}. This also means that the \textit{ForexAPI} knows of most functionality in the system, resulting in unintended functionalities having been implemented directly
in the service over time. The \textit{ForexAPI} runs as a service, in its own process, exposing its interface over both a web-service interface, via messages on \textit{RabbitMQ} and a TCP socket for synchronous \textit{RPC}.

\paragraph{RequestService}
The \textit{RequestService} receives requests for trades and line-checks from \textit{ForexAPI} and feeds them to the mainframe. This includes translation to and from the mainframe, with which it communicates over a single TCP socket. Beyond this, the service also provides data and information from the mainframe and \textit{ForexData} to the \textit{ForexClient}. Most of the business logic lies within this service as well, including authentication of clients and requests, trade responsibility assignment, trade validation, trade registration and line-check processing with data from the mainframe. 

The \textit{RequestService} shares some of its business logic with the \textit{ForexAPI}, for example trade registration logic (processed by both components), and the knowledge of all foreign protocols from the external APIs. Records of all received messages and trades processed are stored in the database \textit{ForexData}. The \textit{RequestService} runs as a service, in its own process, exposing its interface over a TCP socket for synchronous \textit{RPC}.

\paragraph{ForexData and ForexBasicData}
The states of both \textit{ForexAPI} and \textit{RequestService} are persisted in the relational SQL database \textit{ForexData}, which includes records of trades alongside with the original raw messages from the external APIs. The data is also used to integrate some of the trade-registration logic, spread across \textit{ForexAPI} and \textit{RequestService}. The \textit{ForexBasicData} service synchronizes some of the often accessed static data from the mainframe database in order to speed up access, acting therefore like a cache. However, differently from a cache, some data is sometimes also synchronized down to the mainframe from \textit{ForexData}.

\paragraph{PushService}
It listens to updates on trades and line-checks in the mainframe and fetches additional information from the \textit{ForexAPI}.It pushes updates to the \textit{ForexClient}.

\paragraph{Shared Libraries}
Following the \textit{Do not Repeat Yourself (DRY)} principle \cite{DRY}, a number of shared libraries and components have been created, which are used across the system. These can be seen in the upper left corner, in Figure~\ref{fig:danskebank_monolith}. They are simple \textit{.NET DLL's}, which are maintained and used across almost all components of the system and include a unified model in \textit{Models} and access to the database through \textit{DataAccess}. The libraries have dependencies between each other, resulting in difficulties when it comes to updates.

\paragraph{Mainframe and DB2}
Although the mainframe and its associated database \textit{DB2} are not
official parts of the \textit{FX Core}, the system relies on its functionalities, such as fetching of account balances used for line-checks, and the final registration of trades, i.e.\ requests to move assets from one account to another. It also contains organization information, such as users and their access rights, which is used for authorization purposes. The integration with the mainframe happens via \textit{RPC} calls over a TCP socket.

\subsubsection{Integration}
\label{subsubsec:fxcoremonolithintegration}

A wide variety of integration mechanisms and technologies are used between
components and to external clients. In the following we provide a brief description.\\

\begin{itemize}
\item \emph{Proprietary external protocols} from external clients and providers of trade and line-check requests, to the \textit{external APIs}. Protocol messages are sent to the system through a TCP socket established between the providers and the \textit{external APIs}. One of these proprietary protocols is the \textit{FIX} protocol, which is used by many financial institutions.

\item \emph{.NET RPC} over TCP is used to integrate some of the internal components, \textit{RequestService}, \textit{ForexAPI} and \textit{PushService}.

\item \emph{Messages via RabbitMQ} is used to integrate the \textit{external APIs} with the \textit{ForexAPI}.
  
\item \emph{Web-service interface} in the form of \textit{Windows Communication Foundation (WCF)} and \textit{SOAP}, provided by the \textit{ForexAPI} to some clients, including traders wishing to manually fetch information.

\item \emph{Mainframe calls} are done over a proprietary RPC protocol on TCP sockets, and is used by most of the services in the system, to integrate with functionality and data in the mainframe.
  
\item \emph{Database integration} is used between \textit{ForexAPI} and \textit{RequestService} for some functionalities, such as trade-registration, meaning that instead of communicating trade registration data directly, they do it indirectly through the database. It is also used by some traders and other external systems to fetch data directly from the database.
\end{itemize}

\subsubsection{Deployment}
\label{subsubsec:fxcoremonolithdeployment}

\begin{figure}[!htbp]
    \centering
    \includegraphics[width=0.5\textwidth]{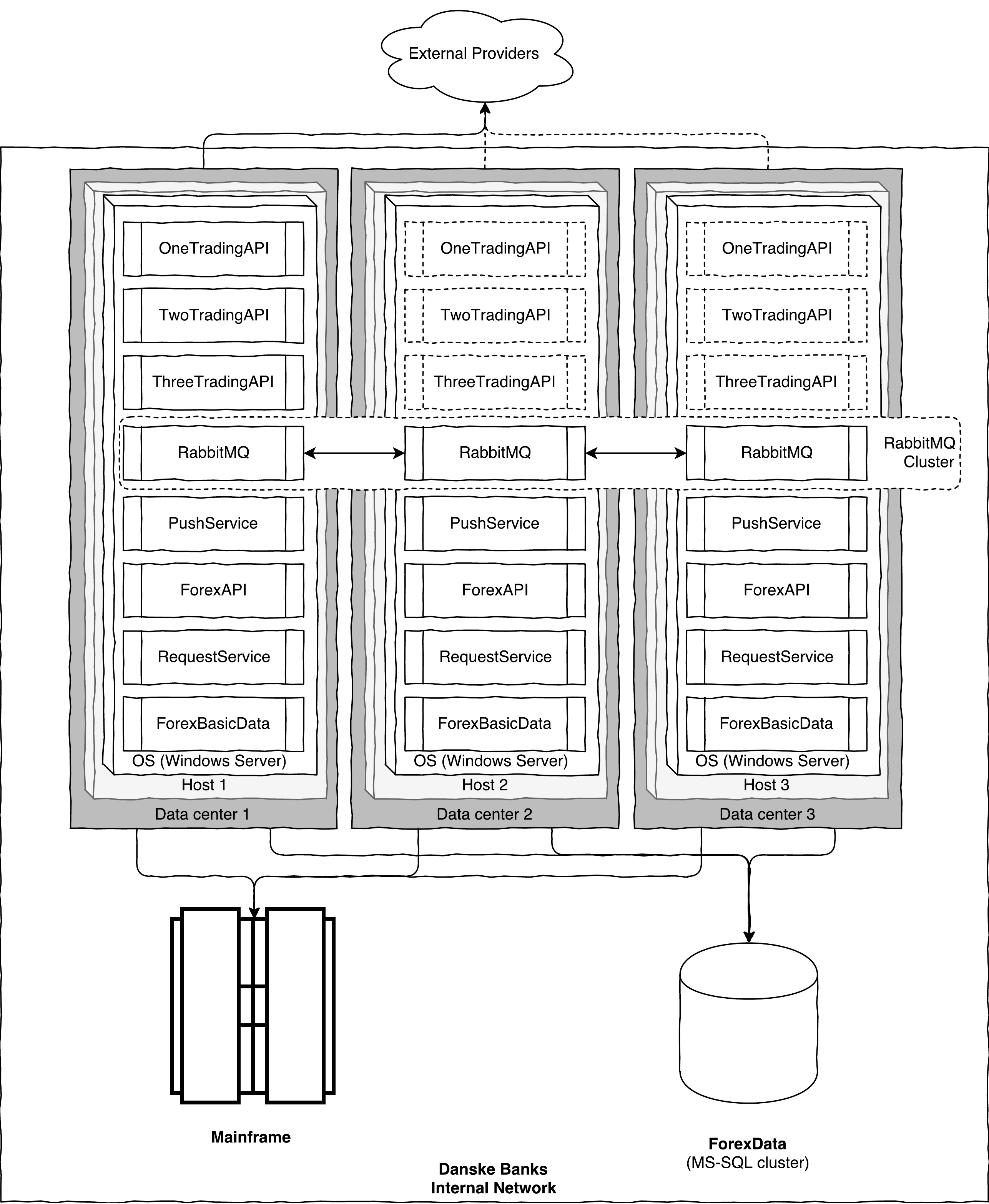}
    \caption{Illustration of how the older monolithic architecture is
    deployed on Danske Banks internal datacenters, which are connected via
    a \textit{VPN} on a \textit{WAN}. Three servers are provisioned
    to run the system, and the whole systems is replicated across the three
    servers. The dashed references indicate fail-over connections and dashed
    processes indicate fail-over replicas running in active/passive. Meaning
    that, should the active (solid) fail, the passive (dashed) will take over. 
    The mainframe and ForexData database are not deployed together with the
    system, but are managed by the IT department. The external providers are
    located outside Danske Banks internal network. All components in the system, run directly on the OS, which in this case is Windows Server}
\label{fig:monolith_deployment}
\end{figure}

The system is deployed on three Windows Server hosts, located at the three Danske Banks data center locations, as shown in Figure~\ref{fig:monolith_deployment}. Each of the system's components can be deployed individually, as they are independent processes, but in fact they are always co-located as a whole system for availability reasons and for the components to be highly coupled. This also means that all but the external provider API's are running in \textit{active/active} fail-over, meaning that they all run concurrently, ideally communicating with the instances they are co-located with. Since the team has no extensive monitoring, they can not guarantee that they actually communicate with co-located services, since fail-over will result in services communicating with dependencies on the other servers, and fall-back is not implemented.
All the components hold local references to all instances (replicas) of
the services, on which they depend. This means that in the case a co-located
dependency should fail, they can fail-over and establish a connection to one
of the other instances on another host instead. The external provider APIs all run in \textit{active/passive} fail-over, since only a single socket can be kept open per external provider. If an external provider API terminates, its connection to a provider will be taken over by one of the \textit{passive} replicated instances, hereby becoming \textit{active}.
The problem with both types of fail-overs are that once fail-over has occurred and the failed component is alive again, the dependants will not fall back. This can result in only a single instance actually being active and serving the system, should two previous nodes have failed, whether or not they are alive again. Manual intervention is required to fall back services to their co-located dependency-replica.

The \textit{RabbitMQ} messaging system runs clustered across all nodes, since it is responsible for routing messages. Effectively functioning as a load balancer of trade and line-check requests from external providers. The clustering ensures that no messages from external providers are missed, should a \textit{RabbitMQ} node terminate. Thus maintaining the \textit{at least once} delivery guarantee provided by \textit{RabbitMQ}. All servers are manually maintained, hereby becoming snowflakes i.e.\ manually configured risking heterogeneous and non-replicable environments~\cite{fowler2012a}. Deployment of components is automated with the continuous integration system \textit{GoCD}~\cite{gocd2017}. 

\subsection{Scalability}
The monolithic architecture itself applies scalability techniques that will be described in the following.


\subsubsection{Horizontal Scaling}
The system is scaled horizontally by usage of multiple hosts, with the whole system being deployed across these, as seen in Figure~\ref{fig:monolith_deployment}. This results in higher availability, since redundancy is introduced. Redundancy in the form of multiple
instances running and reliance on fail-over. The implementation of horizontal scaling is not elastic, since all hosts are manually configured and all services need to be configured upon introduction of additional replicas. This is because they hold references to their dependencies and
all dependency replicas internally. This makes it cumbersome to introduce additional hosts, although possible, it is usually not done for temporary scaling needs. This also means that resources are statically allocated, to the maximum possible load.

\subsubsection{Distribution}
Distribution of functionality into services has been implemented. The
implemented distribution does not result in \textit{high cohesion} and
\textit{low coupling}, since the components are implemented with a high degree of coupling. This results in small degrees of load distribution and does not significantly improve load scalability. The usage of functionality distribution into thick clients is utilized, meaning that clients handle a lot of logic for data inspection, e.g.\ filtering,
sorting and visualization, which relieves the monolithic system from some load. Using thick clients should in theory also enable better geographical scalability, as it reduces the amount of networked communication necessary between the \textit{ForexClient} and \textit{FXCore}, thus also minimizing latency. This is not the case with the \textit{ForexClient} though, since it is highly coupled to \textit{RequestService}, resulting in extensive communication between the two.

\subsubsection{Replication}
The whole system, except the external APIs, is replicated and runs concurrently in \textit{active/active} mode across the three servers. The external APIs are replicated, but in \textit{active/passive}, meaning that only a single replica instance is active at a time, see Figure~\ref{fig:monolith_deployment}. This is due to limits on how many sockets are available to external providers. This replication allows load to be split amongst replicas, resulting in better load scalability. The replicas also act as redundant instances hereby improving availability.

\subsubsection{Concurrency}
Since all trades and line-checks are independent they can be executed
concurrently. The trades are spread amongst the replicated systems by
\textit{RabbitMQ}, meaning that multiple trades can be executed in parallel. The services, such as the \textit{RequestService}, also make use of multi-threading, in order to concurrently processes requests.

\subsubsection{Clustering}
In the internal system clustering is only applied to the \textit{RabbitMQ}
messaging, which hereby enables higher availability and increased throughput. Should a \textit{RabbitMQ} node terminate or become unavailable due to a network partition, the cluster will automatically handle partitioning based on its consistency configuration. The \textit{ForexData} database is stored on a database cluster, but this is
managed by the IT department and is therefore not the teams responsibility to make it scalable.

\subsubsection{Caching}
The \textit{ForexBasicData} component mirrors some data to \textit{ForexData} from the mainframes DB2. This is somewhat a cache, as it speeds up access to some static data from the mainframe, but it is not significantly faster as the database is not optimized for fast reads. Additionally many of the services use simple internal memory caches, mostly caching data from database and mainframe, in order to reduce latency on serving requests.

\subsubsection{CAP Theorem}
The CAP theorem \cite{Brewer2012}, also known as Brewer's theorem, states that it is impossible for a distributed computer system to simultaneously provide all three of the following guarantees: Consistency, Availability, and Partition tolerance. In network partition scenarios only the clustered components of the system are required to choose between consistency and availability., i.e.,\ only \textit{RabbitMQ} since \textit{ForexData} is an external dependency. Since availability is both a goal with scaling the system and a desired characteristic, the system is optimized for this. \textit{RabbitMQ} is configured to handle partitions with its \textit{auto-heal} feature, which is optimized for availability during network partitions~\cite{rabbitmq2016partitions}.

\subsubsection{Fault Tolerance}
Fault tolerance is mainly implemented as part of the fail-over mechanisms in replication, load-balancing and routing. This ensures that if a component of the system fails, a replica is ready to take over its load.

\subsubsection{Load Balancing and Routing}
Load balancing is mainly handled by \textit{RabbitMQ} which will distribute messages from the external APIs in a \textit{round-robin} manner between replicas of \textit{ForexAPI}. For other users of the system, e.g.\ the \textit{ForexClient}, load will be determined by which host they have a reference to, which can be configured on the clients from a central configuration management tool. Routing between the system components are manually configured, so each instance of a component has a list of all available replicated instances of their dependency. The co-located instance of a dependency will mostly, sometimes not, be chosen first, but should that terminate or become unavailable, fail-over to an instance on another server will happen.This fail-over is somewhat
faulty, since if the failed dependency becomes available again, routing will not rollback to the original instance. The load balancing improves load scalability and availability, as it enables splitting the load across multiple replicas. The internal routing also ensures that a component stays available, should a dependency fail. The routing can also be used to route requests to the replica nearest the requester, hereby improving geographical scalability and latency.

\subsubsection{Health-checking, Monitoring and Logging}
The monolithic architecture has no centralized logs or monitoring, but instead relies on manual inspection and reports of erroneous behavior from the users. All system components create local logs, which are manually aggregated, searched and investigated by the developers, as a \textit{post-mortem} analysis, i.e.after errors have occurred and are observed. The only preventive monitoring, is done by inspecting the size of and errors in the error queue, within \textit{RabbitMQ}. This is where messages are located, when they have not been handled successfully. This is continuously monitored on a screen within the office.

\subsection{Achieved Characteristics and Goals}
Here we will provide a summary of the (fully or partially) achieved scalability characteristics.

\begin{itemize} 
\item \emph{Load scalability} has been achieved since it can handle load up to the statically allocated resources limits. The system can be expanded manually if more load should occur.
    
\item \emph{Geographical scalability} has been achieved to a limited degree. Since the system is accessed by external providers and the system is not latency critical, geographical scalability is mostly a question of keeping the system available via the Internet. Since components are highly coupled, the integration between them is extensively chatty, thus also reducing geographical scalability.
    
\item \emph{Elasticity} has not been achieved, since the infrastructure cannot expand and contract based on load. Additionally the architecture is not suited for dynamic additions of resources and replicas, as load balancing is mainly done by the requester having references to all replicated dependencies.
    
\item \emph{Fault-tolerance} is achieved to a limited degree, since
 it can handle faults with fail-over, but fall-back after a failed component is alive again does not function optimally.

\item \emph{High Availability} is achieved to a limited degree, since the system has implemented fault-tolerance mechanisms in the form of redundant replicas, fail-over and configuration of \textit{RabbitMQ} to prefer availability during network partitions. But since the system does not provide any centralized aggregation of health-checking, monitoring or
logging, the system or developers can not act in a preventive manner, which
could lead to reduced availability.
    
\item \emph{Weak Consistency} is achieved by configuring \textit{RabbitMQ} to prefer \textit{availability} over \textit{strong consistency}. \textit{ForexData} is outside of the system and the database is a transactional MS SQL database, which ensures that writes are \textit{strongly consistent}. The asynchronous \textit{at-least-once}    delivery guarantee and the \textit{strong consistency} guarantee of MS    SQL, results in the  system becoming \textit{eventually consistent}.
\end{itemize}

Below is a summary of which techniques have contributed to the scalability goals.

\begin{itemize}
\item \emph{Throughput} has been improved by implementing horizontal scaling, replication, concurrency, and load-balancing. Distribution of functionality has not contributed much to throughput, as the components are highly coupled.
    
\item \emph{Availability} has been improved by implementing horizontal scaling, replication, load-balancing, clustering and fault-tolerance. There is still room for improvement, by fixing the fail-over mechanisms and by allowing the system or developers to act preventively on aggregated health-checking, monitoring and logging. The CAP theorem has also enabled the team to configure clustering to prefer availability.

\item \emph{Latency} has been improved by caching, load-balancing, replication and concurrency. But the tight coupling between services and the use of many different communication paradigms makes it difficult to improve and optimize even further. Distribution might also have introduced some extra latency, since the system can no longer rely on IPC for communication and since services are tightly coupled the chatty communication between them might introduce even more latency.
\end{itemize}

\subsection{Problems}
Beyond scalability, the system has some other problems which has motivated
the team to design and implement a new architecture from scratch.
Below are some of the major problems with the old monolithic architecture.

\subsubsection{Large Components}
As many other organizations experience when growing a system and expanding its functionality, at some point the systems components grow too big. This system suffers from individual monolithic services, which contain too much
functionality to easily comprehend by the engineers, resulting in unnecessary
complexity and confusion on where to locate new functionality which hinders
development. A service such as the \textit{RequestService} suffers from size and contains huge amounts of functionality and complexity and even shares some of its functionality with \textit{ForexAPI}. As visible in Figure~\ref{fig:danskebank_monolith}, it interacts with nearly all other
components in the system, making it both a critical and complex component to
handle. The size and comprehensiveness of the components makes it difficult to know which functionality fits where and has over time resulted in lower cohesion and higher coupling, especially between \textit{RequestService} and
\textit{ForexAPI}.

\subsubsection{Shared Components}
Although the system is split into separate services a lot of functionality is
shared across them in the form of shared components, as it can be seen in
Figure~\ref{fig:danskebank_monolith}. Since the components are shared across
the services, an update of a shared component can result in requiring updates all services. This requires comprehensive testing of the changes across all dependants of an updated shared component. It also results in higher coupling between services, since all services now need to adhere to the standards within the shared components. Low cohesion is also an effect, since the functionality of shared components are now shared amongst dependant services, resulting in functionality associated with the shared component being spread across the multiple dependant components. It is simply put, too convenient to use shared components, which results in unclear boundaries of functionality between services, and results in unnecessary coupling.

\subsubsection{The Mainframe}
Due to the age of the system a lot of the business logic and data is located
within the organisations mainframe. The developers of the system estimate that around 90\% of the business logic is still located in the mainframe. This of course results in some difficulties. Firstly the code on the mainframe is near impossible to comprehend by any of the developers due to it's structural complexity. Most of the code is just imperative and contains no object oriented paradigms or the like. This results in calls and dependencies criss-crossing the system, with no kind of management or overview, making it  extremely difficult to modify or expand on the system. Furthermore, the system  is developed with old legacy technologies, such as \textit{Cobol} and  \textit{DB2}, which slowly are forgotten as a new generation of employees substitute the previous experts on these. The old technologies also make tasks, such as database queries unnecessarily slow, and since it is not trivial to modify the system, optimizations are rarely an option. The mainframe is not an easy component to replace, as it contains much functionality which is core to the business. However, the system owners have started to pull out some functionality into new services and hereby slowly abstracting the mainframe away, in this way minimizing its necessity over time. The \textit{FX Core} monolithic architecture was the first attempt at decoupling the mainframe.

\subsubsection{Complex Deployment}
Deployment is somewhat risky and a very intricate process. Although the system has automated pipelines, the high coupling between components and the usage of shared components, makes deployment pipelines coupled and complex. Updating a single component can result in the whole system requiring a rebuild and redeployment, meaning that the whole system now needs to be tested.

\subsubsection{Organisational Culture and Unknown Dependants}
Danske Bank is a huge organisation and the system has a large number of users
outside the IT department, some of which have the capabilities to develop their own solutions, which are dependent on internal system components. An example is business people which rely on the data found in \textit{ForexData}.
Since these people are not educated in software architecture, they have made
the mistake of writing quick scripts which read data directly from the database. This results in difficulties for the team in modifying the database structure, since it might break important business processes they are not aware of. The team has already developed API's for stakeholders and clients to use, but the transition to these is a slow process. At some point the system needs clear boundaries, hindering practices such as these which slows development and can cause unknown errors.

\subsubsection{Multiple Communication and Integration Paradigms}
The system utilizes many different integration and communication paradigms. This makes communication with the architecture and integration between the components unnecessarily complex, often resulting in violation and bad definitions of interfaces. The usages of \textit{RPC} and \textit{database integration} also results on higher coupling between the components.
Services also communicate directly and often two-way, resulting in even higher coupling.

\subsubsection{Technology Dependence}
Having software in monoliths, also limit the use of technologies. If a developer starts developing a new significant feature within the monolith, said developer is limited to the technologies the monolith is already implemented in, although another technology might be a better fit for the feature and the developers expertise. New technologies come along all the time and for a company to stay relevant in this market of rapidly moving technologies, they must be able to apply some of them in their system.
Although the system is not one big monolith, the choice of databases, integration paradigms, reliance on shared components and choice of deployment platform, limits the usage of technologies to Microsofts \textit{.NET} platform. The other way around, the heavy reliance on Microsoft technologies also limits the deployment platform to \textit{Windows Server}, which in general provides less flexibility for the team in choice of new technologies, automation of infrastructure and the like, compared to running on \textit{Linux} servers.

\subsubsection{Missing System Status Overview}
Since the system does not have a central location and aggregation of monitoring data, health-checks and logs, there is no way to get an overview of the systems status. This missing overview and aggregation requires developers to manually investigate logs or the messaging system for
errors. This means that the developers have minimal opportunity to apply
preventive measures based on warning signs from the system, and mostly rely on \textit{post-mortem} analysis, after the error already has occurred.

\section{FX Core Microservice Architecture}
\label{sec:fxcoremsa}
\begin{figure}[!htbp]
    \centering
    \includegraphics[width=0.5\textwidth]{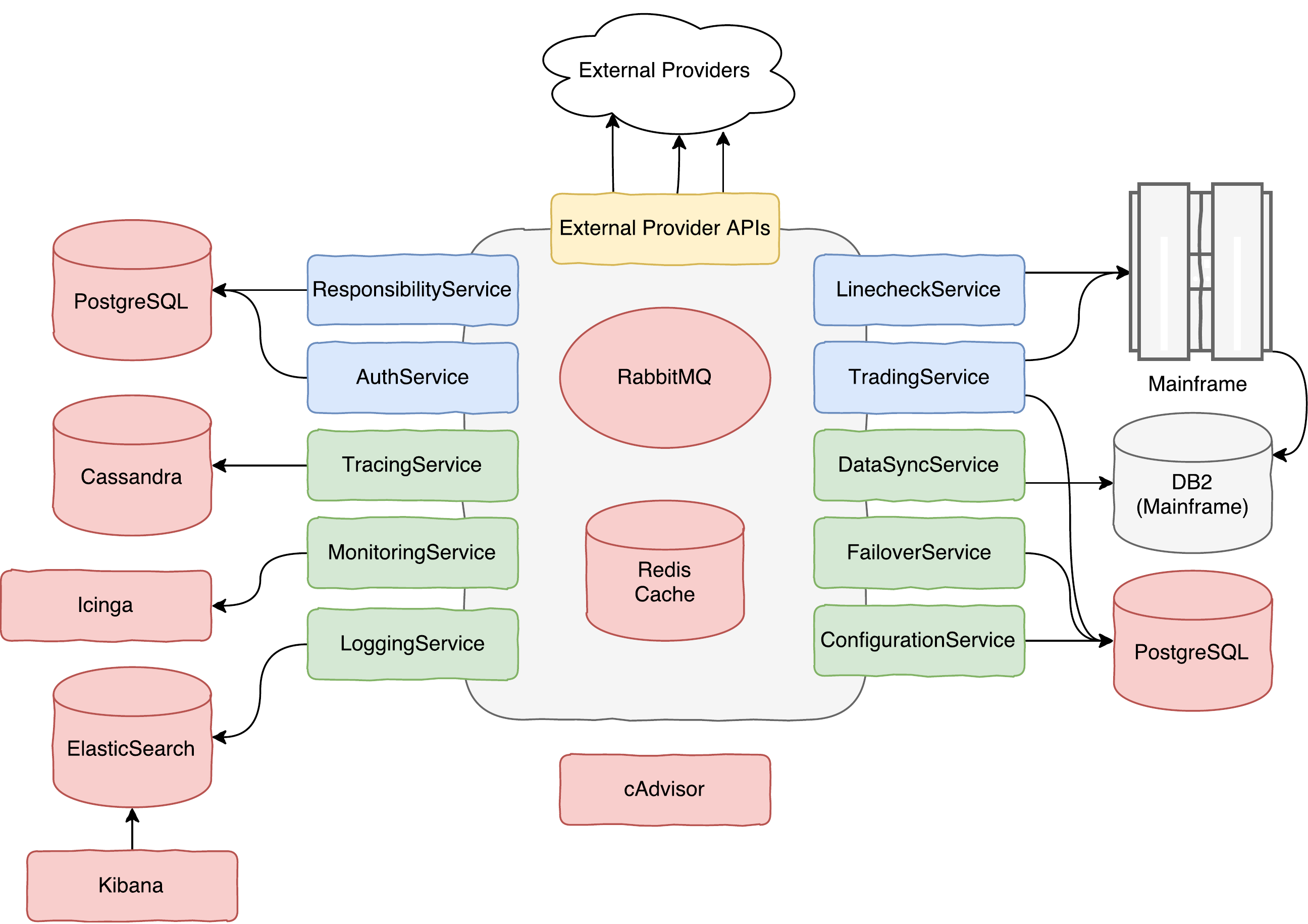}
    \caption{The new \textit{FX Core} microservice architecture. Red services
    are infrastructure services, green are foundation services, blue er business
    services and the yellow is external provider APIs. All non-infrastructure
    services communicate via messaging over \textit{RabbitMQ} and have direct
    access to the \textit{Redis cache}, which is used to cache data from
    \textit{DB2}. Databases in the diagram should be seen as database management
    systems (DBMS), meaning that although four services use \textit{PostgreSQL}
    they all have  their own standalone database within the DBMS}
\label{fig:danskebank_newarchitecture}
\end{figure}

The Danske Banks new \textit{FXCore} architecture is based on the microservice architectural style and is is intended to completely replace the old monolithic architecture. This section will cover how Danske Banks \textit{FX Core} team has chosen to implement a microservice architecture, thus giving an idea of how such an architecture can be implemented in an enterprise setting. This includes a description of the infrastructure as depicted in Figure~\ref{fig:danskebank_newarchitecture}, a brief description of the implemented services, some of the additional architectural principles used, what scalability techniques have been applied and what scalability characteristics and goals have been achieved.

Danske Banks \textit{FX Core} microservice architecture is hosted on private data-centers, i.e.\ not in the cloud. This means that new hosts can not be provisioned and de-provisioned as rapidly and automated as in a cloud. It is in their interest to provide a private cloud for systems to run in, but due to regulations on banking data, this is still work in progress. There are three data-center locations in Denmark, which can be utilized to achieve better availability and increased resilience to their internal systems.

On the IT departments roadmap is the adoption of the \textit{Red Hat
OpenShift}~\cite{openshift2017} Iaas/PaaS platform, on the internal data-centers. However, at the moment, the infrastructure consists of VM's ordered through a web-portal, and which are setup manually by the \textit{FX Core} team.

\subsection{Containerization}
All services in Danske Banks \textit{FXCore} architecture are hosted in
\textit{Linux Containers} on the \textit{Docker Swarm} cluster~\cite{dockerswarm2016}. Containerization enables a whole suite of tooling, provided by the \textit{Docker} platform. \textit{Docker Compose}, for example, allows the whole architecture to be deployed with a single command so that developers define all service dependencies to a service and deploy them for local testing during development. Services are deployed locally, since they are running in containers, but their environments are exactly as if deployed to production.

All container images are hosted on an internal \textit{Docker Registry}, i.e.\
a central repository for container images, with \textit{Dockers} official
registry being \texttt{hub.docker.com}~\cite{dockerhub2016web}. New images are
deployed to the internal registry when a new version of a service is successfully built and tested by the \textit{continuous integration} system. Furthermore, all infrastructure service images and base images from which the services images inherit, are also located on the hub. A list of all FX Core images can be retrieved with a search to the local registry.

\subsection{Automation}
All services in the architecture, including infrastructural clusters, has an
automated \textit{continuous integration and continuous deployment (CICD)} pipeline on their internally hosted \textit{GoCD} server~\cite{gocd2017}.
The \textit{GoCD} platform offers a simple interface, which gives an overview
and interaction with building, testing and deployment. The tooling which comes with the orchestration system \textit{Docker Swarm}, has API's which enables automation of many infrastructural tasks, such as rolling updates. These are utilized by the CICD system, combined with checks on correct functioning.

\subsection{Orchestration}
\label{subsubsec:orchestration}
All deployment and execution of services, in containers, is managed by
\textit{Docker Swarms} orchestration on the \textit{swarm} cluster.
\textit{Swarm} uses the notion of a \texttt{service} which is an aggregation of containers, meaning that multiple replicas of the service containers are treated as a single swarm \texttt{service}, also called \textit{managed containers}. An example of this can be seen in Figure~\ref{fig:msaloadbalancing}, where the
service \texttt{trading-service} has multiple replicated containers, i.e.\
\texttt{trading-service.1} and \texttt{trading-service.2}, but service
discovery and persistence in the same database, allows them to act as a single service.The swarm cluster is also managed by \textit{Swarm} and hosts all services on the cluster. The orchestration tooling also handles service discovery and load balancing, and has web and command line interfaces which can be used for automation of \textit{rolling updates}, \textit{scaling} etc.

\subsection{Clustering}
\label{subsubsec:msaclustering}
Clustering is one of the primary techniques used in the \textit{FX Core}. The architecture runs across five virtual hosts located in the three data-centers. On the hosts a \textit{Docker Swarm} cluster has been setup, with each host acting as a \textit{Swarm Node}. This allows the three container engines on the \textit{Swarm Nodes}, i.e.\ \textit{Docker Engines}, to act as a single engine, allowing containers to run spread across the cluster. This is illustrated in Figure~\ref{fig:msaloadbalancing} Since \textit{Swarm} is also a container orchestration system, it provides the features mentioned in Section~\ref{subsec:enablingscalabilityorchestration}.

\textit{Docker Swarm}~\cite{dockerswarm2016} allows for overlay networking, which enables the developers to define internal networks which is used to communicate between service containers, which all expose their ports to the internal network. \textit{Docker Swarm} also allows for management of storage volumes, which are spread across the cluster nodes and are used to persist data from databases and \textit{RabbitMQ}. The cluster allows for dynamic joining and leaving of \textit{Swarm nodes}, and automatically rebalances location of services to efficiently use resources.

Beyond clustering the container engines, some of the services also utilize
clustering. This includes the messaging system \textit{RabbitMQ}, monitoring
system \textit{Icinga} and all databases, i.e.\ \textit{Redis},
\textit{Cassandra} and \textit{PostgreSQL}. These services
use clustering mostly due to requirements to their availability, since they are
critical components of the infrastructure. Therefore all infrastructure service
clusters are deployed with a service cluster node on at least one Swarm cluster
node in each datacenter. Ensuring that the system can keep running as long as a
single datacenter is available and has an active Swarm cluster node.

\begin{figure}[!htbp]
    \centering
    \includegraphics[width=0.5\textwidth]{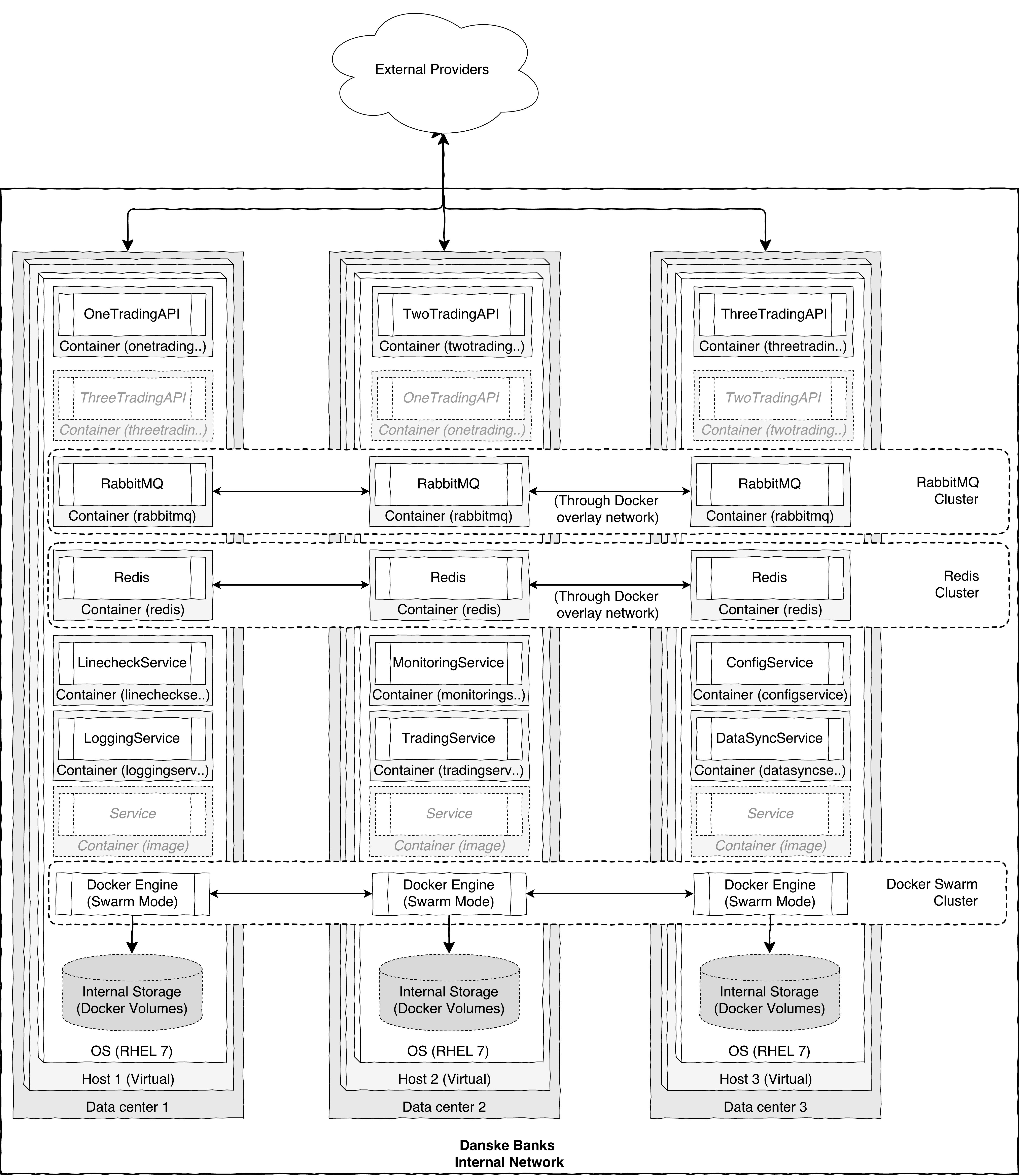}
    \caption{Danske Banks microservice architecture is deployed in three
    datacenters, with at least one host in each datacenter. The amount of nodes
    can vary, but currently five are running spread across the three
    locations. The operating system on all hosts is \textit{Red Hat Enterprise
    Linux (RHEL) 7}, which has been configured to open network ports for the
    installed \textit{Docker Engines} to run and communicate. The \textit{Docker Engines} are
    configured to run as a cluster, i.e.\ each in \textit{Swarm Mode} as part of
    the \textit{Swarm Cluster}. The infrastructure services, such as databases
    and messaging software, here \textit{Redis} and \textit{RabbitMQ}, are
    configured to run on at least one host in each datacenter, and do so in
    clusters. All services can run replicated and be located wherever in the
    cluster, resulting in a heterogeneous deployment. Services running in
    \textit{active/passive} failover, such as the \textit{TradingAPI} services here, also
    run replicated across the cluster}
\label{fig:danskebankclustering}
\end{figure}

\subsection{Load Balancing and Service Discovery}
\label{subsubsec:msaloadbalancingservicediscovery}
Service discovery is implemented as part of \textit{Swarm}, which ensures that
service hostname lookups from containers are translated into IP's of concrete
containers.  Since \textit{RabbitMQ} is used to communicate between services, Swarms's service discovery is only used by services requesting infrastructure services. \textit{RabbitMQ} knows of services which have actively subscribed to one of it's queues, hereby not needing service discovery. Load balancing is therefore required to be implemented by both \textit{RabbitMQ} and \textit{Swarm's} service discovery.

\textit{RabbitMQ} implements load balancing by distributing messages between
subscribers to a queue, hereby spreading load between them. Usually all replicas of a service subscribe to the same queue, and they can hereby share the load. This is usually done in a \textit{round-robin}, distributing messages to replicas in sequential order. \textit{RabbitMQ} queues rely on acknowledgements upon successful processing of a message. Should a replica not
acknowledge a message, it will simply be handed to the next replica, ensuring the message is processed \textit{at-least-once}. In the case no replica can handle the message, it will be sent to an error queue, hereby notifying the developers. \textit{RabbitMQ} can be configured to distribute messages in other ways if load balancing is not needed, which is done upon creation of a queue.

\textit{Swarm} utilizes the built-in service discovery to balance load between
replicated service containers. When a service hostname is requested,
\textit{Swarm} will translate to an IP of one of the replicas containers. For
now, this is also done in a \textit{round-robin}, but one might consider
translating based on \textit{proximity}, i.e.\ to co-located replicas to reduce
latency, or based on \textit{load}, i.e.\ to least busy replicas to improve
throughput.

\begin{figure}[!htbp]
    \centering
    \includegraphics[width=0.5\textwidth]{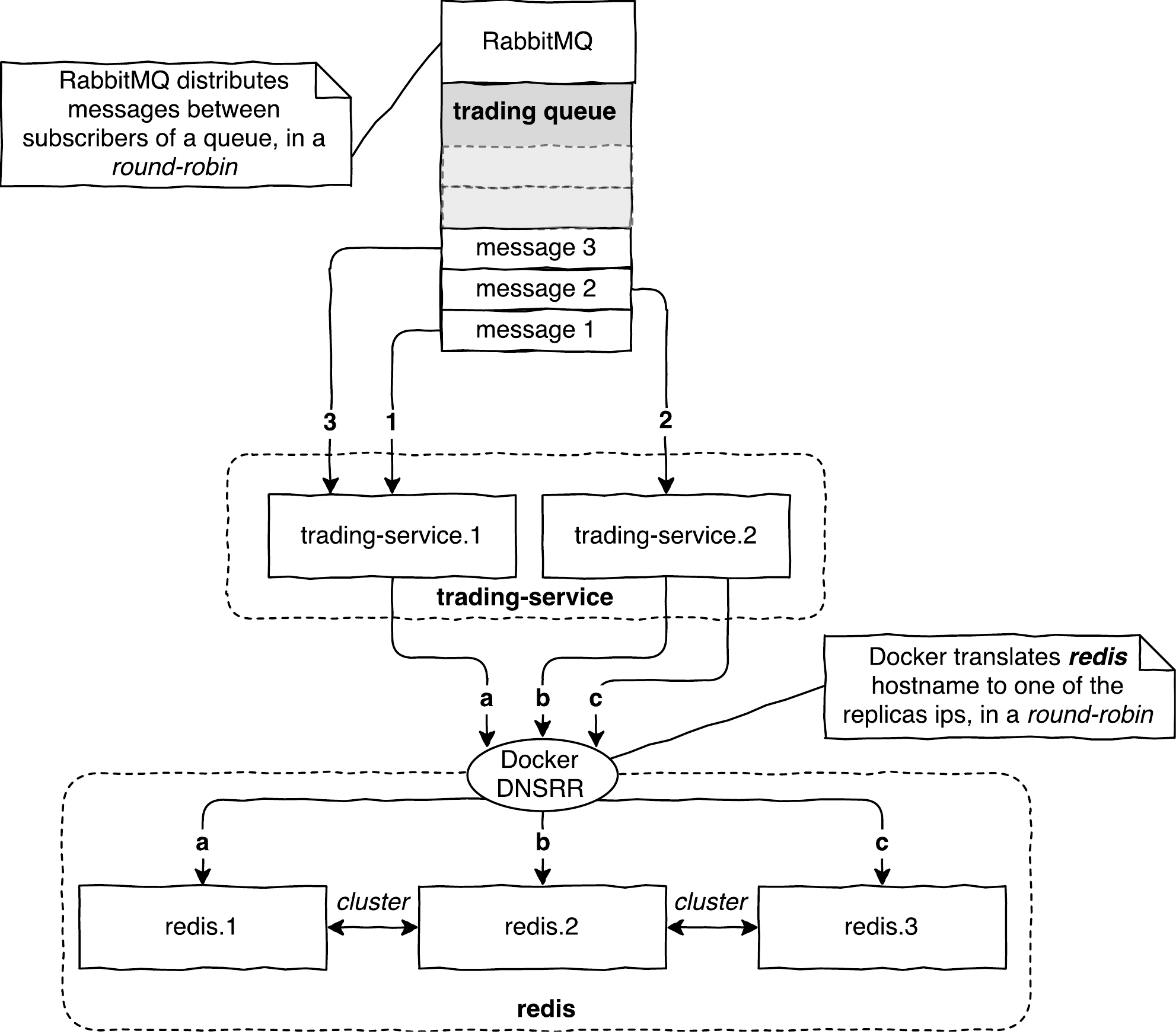}
    \caption{Load balancing is implemented in two places in the infrastructure, as part of \textit{RabbitMQ's} message distribution between replicas and as part of \textit{Docker Swarm's} built-in service discovery.
    \textit{RabbitMQ} distributes messages to replicas subscribed to the same
    queue, and does so in a \textit{round-robin}. \textit{Docker Swarm's}
    built-in service discovery ensures that the lookup of a hostname resolves to a different replica each time, for now also in a \textit{round-robin}. In
    this diagram, when a \texttt{trading-service} replica sends a request to
    \texttt{redis}, \textit{Swarm} will translate the message to one of the
    replicas, e.g.\ \texttt{redis.1}, \texttt{redis.2} or \texttt{redis.3}.
    Since all three replicas run in a cluster, they ensure that state is shared and consistent amongst each other.}
\label{fig:msaloadbalancing}
\end{figure}

An illustration of \textit{RabbitMQ's} and \textit{Swarm's} load balancing, can be found in Figure~\ref{fig:msaloadbalancing}.

\subsection{Services}
\label{subsec:dbmsaservices}
Here will give a short overview of the services within the system, how
they are implemented, how they integrate and their different fail-over modes,
which are important to how they are scaled. The services' responsibilities and
functionality will not be covered in depth, but it should be apparent from their naming.

\subsubsection{Integration}
All services written by Danske Bank integrate via \textit{message-based choreography}. The FX Core team chose message-based choreography because it is asynchronous and decouples services entirely. The chosen messaging system is \textit{RabbitMQ}~\cite{rabbitmq2016}, which provides configurable \textit{publish/subscribe} mechanisms in the form of messaging exchanges, queues and bindings between these. Typically a service which produce messages will do so to an exchange and services which consume messages will do so from a queue. Between exchanges and queues are bindings, which define how messages are distributed from exchange to one or more queues, based on message metadata.
This decouples services from each other and makes all communication between them asynchronous.

Queues are used for specific services and functions as a \textit{load balancer}
between consuming replicas. When a queue is shared among replicas of a service,
they will get supplied with messages in a \textit{round-robin} manner.
This is illustrated in Figure~\ref{fig:msaloadbalancing}.
\textit{RabbitMQ} supports acknowledgements from consumers, so if a message is not
acknowledged after some given timeout it will be redistributed to another
replica, if the redistribution happens too many times, the message will be sent
to an error queue. Acknowledgements are put in place in order to ensure that all
messages are handled eventually and if not the developers will be notified from
the error queue.

\subsubsection{Fail-over Modes}
All the services are categorised into two failover modes, which not only
describes how the services handle failure, but also helps define how they are
run in production.

\begin{itemize}
\item \emph{Active/Active }failover means that multiple service replicas can run alongside each other. This provides them with the ability to share load for better scalability and if one fails, for the others to take over its intended load while it recovers. This can also be utilized for rolling updates, where replicas can be updated one at a time, resulting in zero-downtime updates.

\item \emph{Active/Passive} failover means that only a single instance of a service can be running at a time. Therefore it cannot be scaled by replication, but only by increasing resources, i.e.\ vertical scaling. During runtime a passive service will be idling until the active service fails, and the passive service will become active and take over the workload from the failed service. The same approach applies to updates, where a new version of the service will be deployed and take over the old versions workload when ready, hereby letting the old service terminate.
\end{itemize}

\subsubsection{Foundation Services}
These services function as the foundation of the architecture, meaning that they implement supportive functions and not business related functionalities. They implement centralized logging and monitoring, centralized service configuration and handling of \textit{active/passive} failover. All of these services run in \textit{active/active} failover, meaning they can be replicated and run concurrently:LoggingService, MonitoringService, ConfigurationService, FailoverService, DataSyncService, TracingService.

\subsubsection{Business Services}
These are the services that are actually implementing business logic.
They process trades, line-checks and authorization of actions in the system. This is mainly the group of services which will be expanded before deployment to production. All of these services also run in \textit{active/active} failover, meaning they can also be replicated and run concurrently: LinecheckService, TradingService, ResponsibilityService,
AuthService.

\subsubsection{Infrastructure Services}
All of these services make up the infrastructure of the architecture and
includes messaging, monitoring, logging an databases. All of these
infrastructure services run in clusters, to provide high availability and better performance, i.e.\ load scalability.

\begin{itemize}
\item \emph{Elasticsearch} stores logs and health check data from services.
  \item \emph{Icinga }aggregates, visualizes and inspects monitoring data.
  \item \emph{Kibana} aggregates, searches and inspects logs from all services.
  \item \emph{PostgreSQL} is a database used by most of the services which require
    persistence.
  \item \emph{RabbitMQ} is the main messaging system, used by all services.
  \item \emph{Cassandra} is a database used by the \textit{TracingService}.
  \item \emph{Redis} is used as a cache for static data from the mainframe database.
  \item \emph{cAdvisor} is used to retrieve performance metrics about containers and
    hosts, from the Swarm cluster nodes.
\end{itemize}

\subsubsection{External API Services}
These services provide interfaces to the external providers of \textit{trades}
and \textit{line-checks}. Their main task is to have an open socket to the
providers, translate the messages they receive on proprietary protocols to a standard
format that can be fed to the system, via \textit{RabbitMQ}. They are all
\textit{active/passive} as a provider typically only provides a single socket. The names of the concrete services will not be mentioned, as they are confidential.
\section{Monolith vs. Microservices}
\label{sec:comparisondbmonomsa}
After having presented both architectures, we will now discuss how they differ in handling the effects of scale. Beyond comparing their scalability, this section will also explain how the new microservice architecture copes with the problems we have presented for the monolithic architecture.

\subsection{Effects of Scale}
The two architectures both apply scalability techniques in an effort to achieve scalability characteristics and goals, which ensures they can cope with the effects of scale. In the following, we will discuss how they differ in handling these effects.

\subsubsection{Availability} it is handled better by the microservice architecture, since the monolithic architecture has problems with fall-back after a fail-over. They have both applied techniques to improve availability, but the microservice architectures loose coupling and reliance on replication and load-balancing of individual services has ensured availability will not be affected by scale.
    
\subsubsection{Reliability} it may become an issue at scale since both architectures integrate components with unreliable networked communication. In the microservice architecture, all integration between services rely on 
\textit{RabbitMQ} which can be configured to ensure reliable transfer of messages~\cite{rabbitmq2016partitions}. This may apply to the APIs used to integrate with infrastructure services as well. The simpler integration in the microservice architecture, combined with it's principle of \textit{designing for failure}, could result in better tackling of \textit{reliability} at large scale. Additionally the use of containerized and independent environments of the individual microservices, should also provide the same reliability between local testing and deployment. This is not the case with the monolithic components, which are run directly on the developers machines for testing and server's OS for deployment.

\subsubsection{System Load} it is handled in both architectures by horizontally scaling the hosts and load-balancing between replicas, thus spreading overall system load between hosts. One might also argure that the microservice architecture, although distributed, with it's loose coupling between services, ensures that messaging does not result in too much network traffic. Elasticity also ensures that the microservice architecture can make use of extra allocated resources, which could be used to reduce system load on individual hosts, when needed.
    
\subsubsection{Complexity} it is handled better by the microservice architecture, although more distributed and thus with more moving parts. This is mainly due to it is high cohesion, low coupling, extensive monitoring and logging, and reliance on automation. Which all contribute to reduced complexity, both in structure, separation of responsibilities and deployment. This will likely be kept over time as the architecture is optimized to be evolved by addition of services. The monolithic architecture has high coupling between it is components, which are all very large, making the architecture's structure and integration complex, with responsibilities not clearly separated. Integration patterns are also many, which also contribute to complexity. Although automated, deployment is still a complex process due to shared components and high coupling between components. When deployed the missing centralized logging and monitoring also makes it complex to keep the system running.
   
\subsubsection{Administration} administrative costs are greatly reduced in the microservice architecture as it relies on orchestation tooling, automation and extensive centralized monitoring and logging. The monolithic architecture requires more manual work to keep running at large scale, as deployments are more complex, there's no centralized location for monitoring or logging and the server environments are manually maintained.
    
\subsubsection{Consistency} in both systems it is simply kept \textit{weak}, or more precisely \textit{eventual}. This also ensures that the system can be kept highly available at large scale, although network partitions might occur.
    
\subsubsection{Hetereogenity} it is handled by the microservice architecture due to the use of containerized environments, making the individual services highly portable. This is also substantiated by it's ability to be deployed to heterogeneous infrastructure, i.e.\ differently sized hosts. The monolithic architecture's use of technologies, also requires it to run in a specifically configured and maintained environment, i.e.\ \textit{Windows Server} snowflakes, making it less portable. It is also required to run on homogeneous infrastructure, i.e.\ same sized hosts, since the whole architecture is deployed to each host, thus requiring the same amount of resources.

From the above analysis, it results evident that the \textit{FX Core} microservice architecture in general handles the effects of scale better than the monolithic architecture.

\subsection{Solving Monolithic Problems}
Let us now see how the microservice architecture has improved or solved some of the problems identified in the monolithic architecture. 

\paragraph{Large Components}
The large components of the monolithic architecture which were highly coupled,
had overlapping responsibilities and integrated in a multitude of ways, have been
substituted with several independent microservices.
Just the name of the services reveal their responsibility and they are generally
way smaller compared to the large monolithic services. They do not integrate
directly, resulting in looser coupling and less chance of feature overlapping in
the future. As an example, trade-registration and line-checks were handled both by
\textit{ForexAPI} and \textit{RequestService} amongst almost all other
functionality in the monolithic architecture. In the microservice architecture
a \textit{TradingService} and a \textit{LineCheckService} are handling these
tasks individually instead. This is the case with all other functionalities in
the microservice architecture, resulting in low coupling, high cohesion and
small services.

\paragraph{Shared Components}
The shared component were many in the monolithic architecture, but in the
microservice architecture, this has been reduced to only one shared component,
the \textit{Lambda} framework. \textit{Lambda} is very minimal and is only meant
to be a framework to connect to the infrastructure and provide standard
formatting methods for e.g.\ messages, logs and health-checks.

\paragraph{The Mainframe}
The mainframe will still be attached for some time to come in the microservice
architecture, but over time the functionalities from the mainframe will be
implemented as new services. This will in the future result in all \textit{Forex}
functionality being extracted, totally decoupling the mainframe from the system.
For now, the impact of the mainframe has been reduced by caching.

\paragraph{Complex Deployment}
Since the microservices are independent, loosely coupled and isolated components,
they can be deployed individually, without affecting the other components. There
is no \textit{dependency hell} and the only shared component is \textit{Lambda}.
Even when \textit{Lambda} is updated, all the services are not necessarily
required to update, since they run in their own containerized environments and
do not directly share any dependencies, i.e.\ libraries.
This makes deployment very simple and the usage of Docker and Linux containers
ensures that services run in the same environment during local testing, on test
servers and in production.

\paragraph{Organisational Culture and Unknown Dependants}
The whole re-implementation brings other benefits with it than a new
microservice architecture. It also allows the team to kill all paths into the
system, which they do not control. Since the team controls the whole
infrastructure with Docker, including databases and ports open to outside
clients, the team can eliminate all unwanted access. This allows the team to
develop open APIs for clients and traders in the bank to use, thus eliminating
direct database queries and the like. This gives the team full ownership and
control of internal implementation details.

\paragraph{Multiple Communication and Integration Paradigms}
Internally the microservices integrate only via messaging on \textit{RabbitMQ}.
Due to using \textit{message-based choreography} the services do not call each
other directly, thus resulting in very low coupling and no interfaces to violate.
The system does communicate to external systems via other paradigms, such as the
proprietary protocols to external providers and future \textit{REST} APIs, but
this does not compromise internal system complexity. The integration between
services and their infrastructure dependencies, does not result in internally
complexity either, as it is not used for any integration between business or
foundation services.

\paragraph{Technology Dependence}
The team aimed for a polyglot architecture, meaning that it is not technology dependent. The team is no longer dependent on the \textit{.NET} platform or MS SQL databases, but can implement the services in whatever language they like. One might argue that they are just becoming dependent on other technologies, such as \textit{Docker}, but Linux containers are becoming a standard through the \textit{Open Container Initiative}~\cite{oci2016web}. In general, the team now has more flexibility to choose the technologies they see fit.

\paragraph{Missing System Status Overview}
The microservice architecture has centralized logging in the form of
\textit{LoggingService}, \textit{ElasticSearch} and \textit{Kibana}, allowing
for aggregation of logs from all services. The same applies to
monitoring implemented with the \textit{MonitoringService}, \textit{Icinga} and
\textit{cAdvisor}, allowing for aggregated monitoring of metrics.
Centralizing and aggregating both logs and monitoring, gives the team a complete
system status overview, allowing them to act proactively on suspicious and faulty behaviour.
There is still some work in setting up monitoring, and integrating
\textit{cAdvisor} with \textit{Icinga}, but the foundation has been established.
\section{Conclusion}
\label{sec:conclusion}

In this paper we have analyzed a paradigmatic case study of a mission critical system: the \textit{FX Core} system of Danske Bank. We have investigated both the legacy \textit{monolithic architecture} and the new \textit{microservice architecture}. Both architectures have been documented in terms of their design, implementation, applied scalability techniques and achieved scalability. This has
resulted in a thorough investigation of how to implement a scalable microservice architecture, and how complex the transition can be. 

The re-engineering of the system discussed in this paper led to reduced complexity, lower coupling, higher cohesion and a simplified integration. Comparing the two architectural designs, we have seen how microservices led to better scalability as well as offered solutions to the major problems that where caused by the monolithic realization. Although the comparison did not include quantitative metrics, the implementation of specific techniques has been used as an argument in support of increased scalability.

\bibliographystyle{plain}

\bibliography{Bibliography}

\end{document}